# Timed Concurrent State Machines

Wiktor B. Daszczuk

Institute of Computer Science,
Warsaw University of Technology,
Nowowiejska str.15/19, 00-665 Warsaw, Poland
`wbd@ii.pw.edu.pl`

**Title:** Timed Concurrent State Machines

**Abstract.** Timed Concurrent State Machines are an application of Alur's Timed Automata concept to coincidence-based (rather than interleaving) CSM modeling technique. TCSM support the idea of testing automata, allowing to specify time properties easier than temporal formulas. Also, calculation of a global state space in real-time domain (Region Concurrent State Machines) is defined, allowing to store a verified system in ready-to-verification form, and to multiply it by various testing automata.

**Tytuł:** Współbieżne Maszyny Stanowe z Czasem

**Streszczenie.** Współbieżne Maszyny Stanowe z Czasem TCSM są aplikacją Automatów Czasowych Alura w środowisku koincydencyjnycm Współbieżnych Maszyn Czasowych CSM (w przeciwieństwie do środowisk przeplotowych). TCSM pasują do idei automatów testujących, które pozwalają wyspecyfikować zależności czasowe łatwiej niż poprzez formuły temporalne. Ponadto zdefiniowano sposób wyznaczania globalnej przestrzeni stanów w dziedzinie czasu (Współbieżne Maszyny Stanowe Regionów RCSM), co pozwala przechowywać badany system w postaci gotowej do weryfikacji i mnożyć go przez różne automaty testujące.

**Słowa kluczowe:** weryfikacja modelowa, weryfikacja w czasie rzeczywistym, automaty czasowe

## 1 Introduction

In [1,2], Alur presented a successful idea of introducing real time to the specification of concurrent system. A kind of Büchi automata (with real-valued clocks added) is used. The strength of this method is compositionality, i.e. an automaton representing a concurrent system is compound from individual automata of components. In ICS WUT, another modeling technique called CSM (Concurrent State Machines) is used

[9]. The components of concurrent CSM system, called *automata* for simplicity, differ from Büchi automata in several assumptions, the most important are:
- Transitions are triggered by compositions of signals (the domain of transition function is $2^{\{input\_alphabet\}}$ rather than input alphabet itself); this means parallelism related to input alphabet.
- Transitions in distinct automata are executed in parallel rather than interleaved; this means parallelism in actions.

The main notions of TCSM are defined quite similarly to TA. However, the main extension of TCSM is that the product operation is defined for RCSM (corresponding to Region Automata). This unique feature allows a verification system to store a state space of a system under test in a form of RCSM, called "automaton under test". A system is form of RCSM can be checked against temporal formulas, but we prefer other verification technique, consisting of three phases:
- construction of an RCSM "testing automaton" representing needed (or conversely, unneeded) feature; the automaton may be designed by a user or obtained automatically from other form defining the behavior (for example UML sequence diagram, collaboration diagram or state diagram);
- obtaining a product of the automaton under test with testing automaton;
- reduction of output product using the reduction algorithm presented in [7,11,12,13];
- "safety" features are verified by checking the existence of "error states" in reduced output product;
- "liveness" features are verified by testing output product against stuttering of given states of testing automaton (this technique elaborated by Jerzy Mieściki is a subject of a separate paper under preparation).

The presented verification technique is useful for a designer which does not know a temporal logic, or when it is difficult to express the desired feature in terms of a temporal logic. Also, behavioral conditions may be automatically converted to testing automata from other modeling formalisms.

## 2  Definition of CSM

Before introduction of real time, a timeless version of CSM will be defined.
- **universe** $\underline{U}$ - a countable set of elementary symbols called **signals** $\underline{x} \in \underline{U}$,
- **alphabet** $\underline{A} \subseteq \underline{U}$ (finite subset),
- **atomic Boolean formula** $x$ [1]; a formula $x$ is satisfied if a signal $\underline{x}$ occurs,
- **Boolean formula** $w$ – a sentence in traditional Boolean algebra over atomic Boolean formulas $x \mid \underline{x} \in \underline{U}$ and special symbols *{false, true}*, $\vee$, $\wedge$, $\neg$ [2],

---

[1] The usage of underlined signals and occurrences of signals written in italics highlights the difference between the name of a signal (underlined: $\underline{x}$) and expressing the fact that the signal is being generated (italics: *x*).

- *sig(w)* - set of signals occurring in Boolean formula *w*,
- *W* - set of all Boolean formulas *w*.

We will identify a formula *w* with a Boolean function *f* such that for every set of occurrences of signals the formula *w* is satisfied iff the function *f* gives the value *true*. The values of special Boolean formulas are: $\mathbb{0}$ is always *false*, $\mathbb{1}$ is always *true*.

**CSM automaton** $p \stackrel{\text{def}}{=} < S, form, out, s_{init} >$[3]:
- *S* - finite set of **states** (nodes in a graph representation),
- **transition function** *form*: $S \times S \to W$ (labels of transitions in graph); *form* is assumed total, i.e. defined for every pair $(s,s') \in S \times S$ (formulas $\mathbb{0}$ and $\mathbb{1}$ are valid values of *form*); **void transitions** $(s,s')$ such that $form(s,s') = \mathbb{0}$ are usually skipped in a graph representation,
- **output function** *out*: $S \to 2^{\underline{U}}$ is a function assigning subsets of signals to states (if $\underline{x} \in out(s)$ then we say that the signal $\underline{x}$ is generated by state *s*),
- unique **initial state** $s_{init} \in S$.

The CSM automaton is assumed to be **transition-complete**, i.e. for any $s \in S$ the disjunction of formulas for all pairs $(s,s') \in S \times S$ is *true*.

Alphabets of an automaton *p*:
- **Input alphabet** $\underline{INP(p)} \subseteq \underline{U}$ – set of all signals referred to in *form*'s range.
- **Output alphabet** $\underline{OUT(p)} \subseteq \underline{U}$ – union of all sets *out(s)*.
- **External input alphabet** $\underline{EXT(p)} \subseteq \underline{U}$ – set of input signals coming from the environment; $\underline{EXT(p)} = \underline{INP(p)} - \underline{OUT(p)}$.
- **Total alphabet** $\underline{ALL(p)} \subseteq \underline{U}$ – union of input and output alphabets.

There is no requirement for sets $\underline{INP(p)}$ and $\underline{OUT(p)}$ to be disjoint: a signal may be generated in a state of an automaton and accepted on a transition in the same automaton.

An **output formula** $\omega(s)$ of the state $s \in S$ is a conjunction of affirmation of all signals belonging to *out(s)* and negations of all signals belonging to *OUT(p)-out(s)*.

A state $s' \in S$ is a **successor** of state $s \in S$: $s \ r \ s'$ iff $form(s,s') * \omega(s) \neq \mathbb{0}$. Note that some non-void transitions may lead to states that are not successors due to signals appearing in output formula, for example if for the state *s* and its successors $s_1, s_2$ in automaton *p*: $form(s,s_1)=ab$, $form(s,s_2)=\neg a + \neg b$, $out(s)=\{\underline{a},\underline{b}\}$, $\underline{OUT(p)}=\{\underline{a},\underline{b},\underline{c}\}$, then $s_1$ is a successor of *s* because $form(s,s_1)*\omega(s) = (ab)*(ab*\neg c) = ab*\neg c \neq \mathbb{0}$, and $s_2$ is not a successor of *s* since $form(s,s_2)*\omega(s) = (\neg a + \neg b)*(ab*\neg c) = \mathbb{0}$.

**Reachability relation** (denoted ***R***) is a transitive extension of ***r***.

Let *p* be a CSM automaton, $w \in W$ be a Boolean formula and $\underline{X} \subseteq \underline{ALL(p)}$. We replace in *w* all atomic Boolean formulas *x* referring to signals $\underline{x} \in \underline{X}$ by symbol $\mathbb{1}$ and all atomic Boolean formulas referring to $\underline{y} \in \underline{OUT(p)}-\underline{X}$ by symbol $\mathbb{0}$. The formula *w'* thus obtained is called **reduced** by *X*, denoted $w' = w \backslash\backslash \underline{X}$. For example if $\underline{OUT(p)}=\{\underline{a},\underline{c},\underline{d}\}$ then the formula $w=ab+\neg c+d$ reduced by the set $\{\underline{a},\underline{c}\}$ is $w'=w\backslash\backslash\{a,c\}=\mathbb{1}b+\neg\mathbb{1}+\mathbb{0}=b+\mathbb{0}+\mathbb{0}=b$.

---

[2] For compatibility with the COSMA environment, the conjunction of signals' occurrences is denoted as * (instead of $\wedge$) or no symbol between signals. Disjunction of signals' occurrences is denoted as + (instead of $\vee$). This notation is used by other authors as well, for example[14].

[3] The symbol $\stackrel{\text{def}}{=}$ denotes equality by definition.

The **Reachability Graph (RG)** of a CSM automaton is the automaton restricted to states reachable from $s_{init}$: $RG \stackrel{def}{=} <GS, form, out, s_{init}>$, where: $GS \subseteq S$ is a set of all reachable states: ($GS \stackrel{def}{=} \{ s : s \in S \wedge s_{init}\ R\ s \} \cup \{ s_{init} \}$); formulas on arcs leading out of a given state $s$ are reduced by the set $out(s)$; all void transitions are removed (domain of *form* is restricted to pairs of states belonging to $GS$).

Let $P$ be a finite set of CSM automata

$P \stackrel{def}{=} \{p_i|\ i=1..n,\ p_i = <S_i, form_i, out_i, s_{i,init}>\}$

The CSM automaton $p = <S, form, out, s_{init}>$ is a **product** of CSM automata from $P$ (denoted $p \stackrel{def}{=} \maltese_{pi \in P}\ p_i$ which means $p_1 \maltese p_2 \maltese ... \maltese p_n$) if sets $\underline{OUT(p_i)}$, $i=1..n$ are pairwise disjoint and:
- $S \subseteq \times_{pi \in P}\ S_i$, ($\times_{pi \in P}\ S_i$ denotes Cartesian product $S_1 \times S_2 \times ... \times S_n$); elements of $S$ (composite states) are tuples of the form: $s \stackrel{def}{=}_f (s_{j1}, s_{j2}, ..., s_{jn})$
- $s_{init} \in S$ is a tuple containing $s_{i,init}$ of all component automata $p_i \in P$: $s_{init} \stackrel{def}{=} (s_{1,init}, s_{2,init}, ..., s_{n,init})$
- for any $s \in S$: $out(s) \stackrel{def}{=} \cup_{pi \in P} out_i(s_{ji})$, $s_{ji} \in S_i$
- for any pair $(s,s') \in S \times S$, $s = (s_{j1}, s_{j2}, ... , s_{jn})$; $s' = (s'_{j1}, s'_{j2}, ... , s'_{jn})$: $form(s,s') \stackrel{def}{=} \wedge_{pi \in P} form_i(s_{ji}, s'_{ji})$; $s_{ji}, s'_{ji} \in S_i$     ($\wedge$ denotes conjunction).

Note that for a product $p$ of CSM automata $p = \maltese_{pi \in P}\ p_i$: the sets $\underline{INP(p)}$, $\underline{OUT(p)}$ and $\underline{ALL(p)}$ are unions of sets of component automata, while $\underline{EXT(p)}$ is not ($\underline{EXT(p)} = \underline{INP(p)} - \underline{OUT(p)}$).

The semantics of CSM is defined formally in [7]. A new state of the system is taken from the transition leading from the current state, according do set of signals on input. Single-step semantics, path semantics and fair path semantics are defined. From now on, we deal with fair CSM automata (with fair path semantics).

## 3  Introduction of real time

A real time is added to CSM model similarly to Alur's Timed Automata [1] (TA).
- The "global time" runs for all automata, getting all values in $R_+$.
- Every automaton has a set of clocks, running synchronously with "global time" or being reset on transitions.

However, there are major differences. The main features of TCSM are:
- In TA time elapses in states of an automaton. Because in CSM there is no concept of "staying in a state" other than executing of a self-loop (called an "ear" due to its graphical image), therefore the flow of time is associated with ears. A state with no ear is instantaneous, i.e. it is left as soon as reached. As in TA, only a finite number of instantaneous states may be visited without elapse of time. A time constraint put on an ear is called *invariant* (similarly to TA).
- Because of the "fairness condition" if there is an escape from a cycle in an automaton, if must be followed after a finite number of loops in a cycle. Therefore, a zero-time loop is not a tragedy if the run may diverge from the loop.

- Only an ending (with no escape) strongly connected subgraph containing no ear is invalid. Also, an automaton containing such a subgraph is invalid.
- Due to the manner of multiplication of automata, an ending strongly connected zero-time subgraph may outcome from a product of valid automata.
- Every timeless CSM automaton is transition-complete, i.e. it acts "somehow" in every situation (it never can be "unexecutable", like TA [1]). Formally, a sum of all formulas on arcs outgoing from a state must be $\mathbb{1}$. In TCSM, to keep the transition-completeness, all formulas must sum to $\mathbb{1}$ in every point in time.
- The important difference between TCSM and TA is the manner of communication. In TA, the communication occurs on common letters of alphabet, and the direction of communication (which automaton is a sender and which is a receiver) is specified separately. In TCSM, signals (symbols of alphabet) are generated in states and accepted on transitions, therefore the direction of com-munication is defined strictly. The example of a system of TA (coming from [2]) converted to TCSM system is presented in Fig. 1. Time dependencies in **TRAIN** model physical delays between sensor signals, in **CONTROLLER** model reaction times and in **GATE** model closing and opening time.

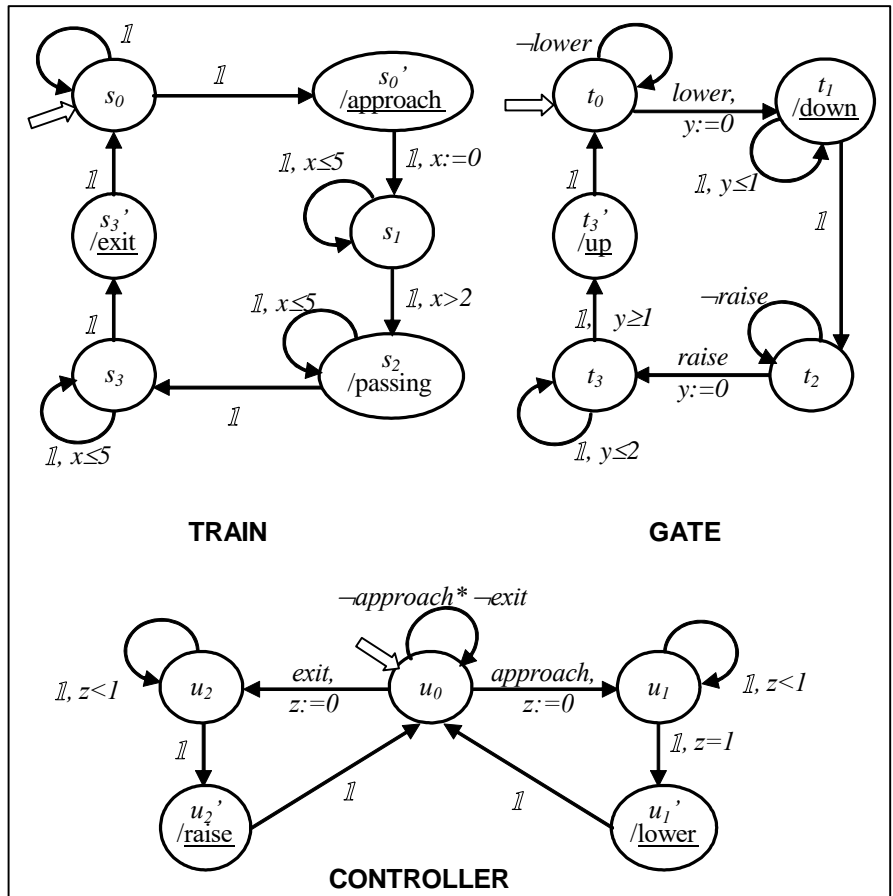

**Fig. 1.** Train-gate controller, Timed CSM

## 4 Definition of TCSM

TCSM automata are based on CSM, therefore only differences will be defined.

Let $X$ be a finite set of clocks (clock variables). Clock constraints are simple constraints $\infty$, $x \leq c$, $c \leq x$, $x<c$, $c<x$ ($c$ is nonnegative real) and Boolean formulas over simple constraints and $\wedge$ (denoted *) The symbol $\infty$ (equals $R_+$) denotes no constraint and may be skipped.

A clock interpretation $\nu$ (from [1]) assigns a real value to every clock in $X$; $\nu$ satisfies constraint $\varphi$ over $X$ iff $\varphi$ evaluates to *true* according to the values given by $\nu$. For $\delta \in R_+$, $\nu+\delta$ denotes the clock interpretation which maps every clock $x$ to the value $\nu(x)+\delta$. For $Y \subseteq X$, $\nu[Y:=0]$ denotes the clock interpretation for $X$ which assigns $0$ to each $x \in Y$, and agrees with $\nu$ over the rest of the clocks.

The **TCSM automaton** is a 5-tuple $p \stackrel{\text{def}}{=} < TS, out, X, lab, s_{init} >$: finite set of **timed states** $TS$ (a shortcut t.state will be used); **output function** $out: TS \rightarrow 2^U$ as in CSM; set of **clock variables** $X$; unique **initial t.state** $s_{init} \in TS$; set of **timed transitions** $lab \subseteq TS \times form \times \Phi(X) \times 2^X \times TS$ (a shortcut t.transition will be used, analogously t.ear).

A t.transition $<s, w, \pi, \lambda, s'>$ from $s \in S$ to $s' \in S$:
- **transition function** $w \in form$ triggers the t.transition (as in CSM);
- **clock constraint** $\pi \in \Phi(X)$ specifies when the t.transition is enabled;
- set of **clocks** to be **reset** $\lambda \subseteq X$.

As a pair *(s,s')* uniquely appoints a t.transition (multiple t.transitions between states are not allowed), elements of a t.transition are extracted using a notation $w(s,s')$, $\pi(s,s')$, $\lambda(s,s')$.

The TCSM automaton is assumed to be **transition-complete**, i.e. for any $s \in S$ and for every clock interpretation, the disjunction of Boolean formulas for all t.transitions outgoing from $s$ is true.

The succession relation cannot be defined for TCSM, because void transitions may occur due to clock interpretations allowed in preceding states (similarly to TA, Region CSM (RCSM), together with succession relation and reachability will be defined in the next section). However, the product of TCSM can be defined.

Let $P$ be a finite set of TCSM automata $P=\{p_i| i=1..n, p_i = <TS_i, out_i, X_i, lab_i, s_{i,init}>\}$.
The TCSM automaton $p = <TS, out, X, lab, s_{init}>$ is a **product** of TCSM automata from $P$ (denot-ed $p \stackrel{def}{=} *_{p_i \in P}\, p_i$ which means $p_1 * p_2 * ... * p_n$) if sets $\underline{OUT(p_i)}$, $i=1..n$ are pairwise disjoint, sets $X_i$, $i=1..n$ are pairwise disjoint and:

- $X \stackrel{def}{=} \cup_{p_i \in P} X_i$,
- $TS \subseteq \times_{p_i \in P} TS_i$;
  elements of $TS$ (composite states) are n-tuples of the form:
  $s \stackrel{def}{=} (s_{j1}, s_{j2}, ..., s_{jn})$
- $s_{init} \in TS$ is a tuple containing $s_{i,init}$ of all component automata $p_i \in P$:
  $s_{init} \stackrel{def}{=} (s_{1,init}, s_{2,init}, ..., s_{n,init})$
- for any $s \in TS$:
  $out(s) \stackrel{def}{=} \cup_{p_i \in P} out_i(s_{ji})$, $s_{ji} \in TS_i$,
- for any t.transition $<s,w,\pi,\lambda,s'> \in lab$, $s = (s_{j1}, s_{j2}, ... , s_{jn})$; $s' = (s'_{j1}, s'_{j2}, ... , s'_{jn})$:
  $w(s,s') \stackrel{def}{=} \wedge_{p_i \in P} w(s_{ji},s'_{ji})$; $s_{ji}, s'_{ji} \in TS_i$.
  $\pi(s,s') \stackrel{def}{=} \wedge_{p_i \in P} \pi(s_{ji},s'_{ji})$; $s_{ji}, s'_{ji} \in TS_i$.
  $\lambda(s,s') \stackrel{def}{=} \cup_{p_i \in P} \lambda(s_{ji},s'_{ji})$; $s_{ji}, s'_{ji} \in TS_i$.

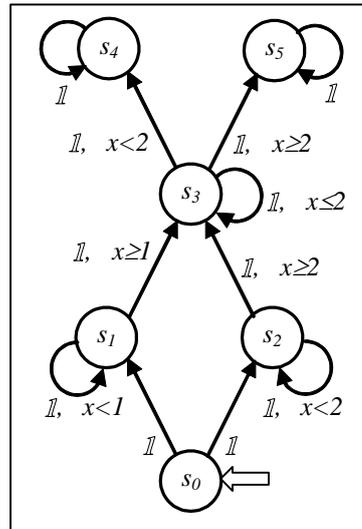

**Fig. 2.** TCSM automaton

The algorithm of obtaining a product of TCSM automata is similar to that for CSM [9] and requires additionally to take on resulting t.transitions conjunctions of time constraints of component t.transitions and sums of clocks to be reset on component t.transitions.

Single step semantics of TCSM can be found in [8]. However, due to lack of reachability relation for t.states, path semantics and fair path semantics cannot be defined for TCSM.

## 5 Region CSM

The single step semantics is not sufficient to express behavior of TCSM automata because of:

- time constraints appearing on transitions (see Fig. 2 showing a transition that is per-

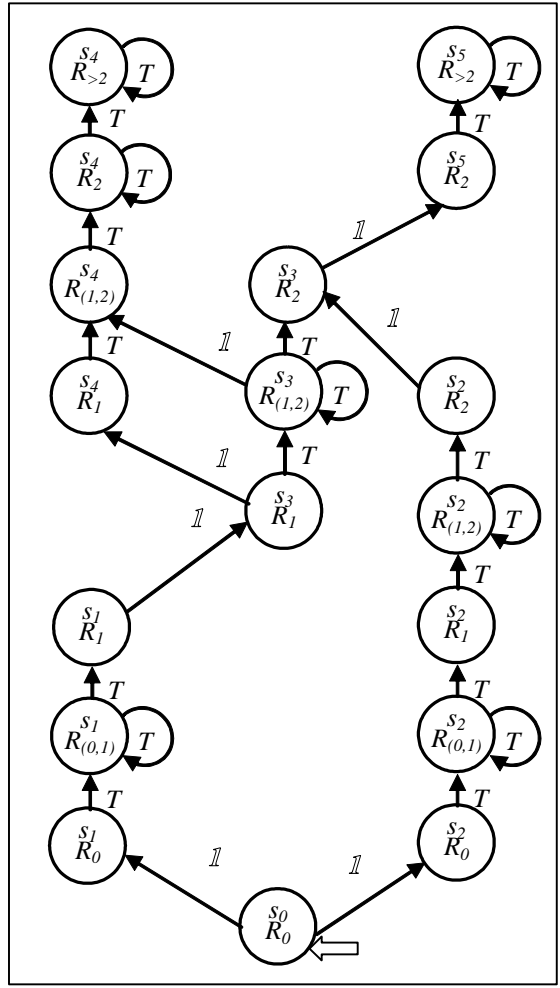

**Fig. 3.** RCSM automaton

sisting in CSM but void in TCSM)
- a fairness condition imposed on the model (Section 2).

Therefore, a Region CSM automaton is defined which allows to observe succession relation and to build a Reachability Graph of a TCSM automaton. It is similar to Region Automaton (RA) of TA [1].

Similarly to RA, constants in time constraints are limited to integral ones (for every TA with real constraints there exists a similar automaton with integral constraints, see [1]). Then, by multiplication of all constants by least common multiple of denominators the set of natural constraints is achieved.

A **timed location** is a pair *(s, v)*. In TCSM, there are infinitely many pairs of timed locations *((s, v),(s', v'))* referring to the same pair of states *(s,s')*. Building RCSM, we achieve finite equivalence classes of these transitions. The regions are sets of clock interpretations that have the same integral part for all clocks and the same fractional part ordering for all clocks. Clock interpretations that exceed a highest constant $c_x$ to which a clock *x* is compared in constraints (clock bound) are not divided into regions.

A region is defined over the set of all clock interpretations for *X*. The automaton of regions where the succession is defined as succession of regions with operations $v+\delta$ and *v[Y:=0]* is stable, i.e. if $v_1$ and $v_2$ belong to common region and some operation ($+\delta$ or *[Y:=0]*) $v_1$ moves $v_1$ to $v_1'$ then there exists a $v_2'$ such that the same operation moves $v_2$ to $v_2'$ and $v_2'$ belongs to the same region as $v_1'$. The stability solves the reachability problems. For example, see Figs. 2 and 3: in TCSM automaton it is not clear if state $s_4$ is reachable from $s_2$ through $s_3$, and in RCSM automaton it is obvious that not (but $s_5$ is). Region states in RCSM are pairs *(s,R)* where *s* is a state of TCSM and *R* is a region. Region index is a range of time interpretations of the clock *x* in the region. Transition marked $1$ are zero-time (non-ear) and transitions marked ***T*** are progress transitions (ear in TCSM, $1$ is skipped).

The proof of region graph stability is given in [2].

Formally [1], for any $\delta \in R_+$, $\lfloor \delta \rfloor$ denotes the fractional part of $\delta$ and $\lceil \delta \rceil$ denotes the integral part of $\delta$, therefore $\delta = \lceil \delta \rceil + \lfloor \delta \rfloor$. For two clock interpretations $v_1$ and $v_2$, they are in common region iff all the following conditions hold (we call them region integrity conditions):

- For all $x \in X$, either $\lceil v_1(x) \rceil$ and $\lceil v_2(x) \rceil$ are the same or both $v_1(x)$ and $v_2(x)$ exceed $c_x$.
- For all $x,y \in X$ with $v_1(x) \leq c_x$ and $v_1(x) \leq c_y$, $\lfloor v_1(x) \rfloor \leq \lfloor v_1(y) \rfloor$ iff $\lfloor v_2(x) \rfloor \leq \lfloor v_2(y) \rfloor$.
- For all $x \in X$ with $v_1(x) \leq c_x$, $\lfloor v_1(x) \rfloor = 0$ iff $\lfloor v_2(x) \rfloor = 0$.

The succession of regions can be easily formulated due to fixed ordering of clock interpretations in a region.

The succession of regions (due to progress of time) results from linear change of values of all clocks (except clocks being reset), and resets of clocks. A region is characterized by a tuple *I* of integral parts of all clocks (or clock bound if a clock interpretation exceeds it), a set of clocks exceeding their bounds, a set of clocks with fractional parts equal to zero, and sets of clocks with equal, non-zero fractional parts; the latter sets ordered from greatest to smallest fractional part. The sets are indexed by > (exceeding bounds), *0* (zero fractional parts), *f1,f2,...* (non-zero fractional parts, or-dered from greatest to smallest). For a set $X$, the sets $Y_>, Y_0, Y_{f1}...F_{fm}$, $m \leq n$, are pairwise disjoint, and they sum to $X$. The rules of succession of regions can be found in [8].

In general, the repertoire of operations over regions must support a succession relation. For example, to the operations defined in TCSM $(+\delta, [Y:=0])$ two additional operations may be added:

- assignment of a natural number (if the greatest number assigned is larger than $c_x$, then $c_x$ must be expanded to this value),
- increment by a natural number ($c_x$ must be enlarged by the greatest number added).

Both these operations preserve region integrity conditions.

An **RCSM automaton** is a 5-tuple $p \stackrel{\text{def}}{=} < RS, out, X, lab, s_{init} >$, where:

- *RS* is a finite set of **region states** *(s,R)*, where s is a TCSM t.state and *R* is a region – a set of time interpretations over *X* (a shortcut r.state will be used for a region state);
- *out*: $RS \to 2^U$ is a function assigning subsets of signals to r.states; the function *out* is called **output function**; if $\underline{x} \in out((s,R))$ then we say that the signal $\underline{x}$ is generated by r.state *(s,R)*;
- *X* – a finite set of clock variables;
- $(s_{init}, R[X:=0]) \in RS$ is a unique **initial r.state**; $(R[X:=0]$ will be denoted $R_{init})$;
- a set of **region transitions** $lab \subseteq RS \times form \times 2^X \times RS$ (a shortcut r.transition will be used); for an r.transition from r.state *(s,R)* to *(s',R')* $<(s,R),w,\lambda,(s',R')>$:
  - $w \in form$ is a Boolean formula triggering the r.transition as before;
  - $\lambda \subseteq X$ is a set of clocks to be reset in this r.transition;
  - for an r.transistion $<(s,R),w,\lambda,(s',R')>$ either *s'=s* and *R'=(R+δ)[λ:=0]* or *R'=R[λ:=0]*.

An r.transition with *s=s'* is called progress r.transition, and an r.transition with $s \neq s'$ is called zero-time r.transition or action r.transition. As a pair *((s,R),(s',R'))* uniquely appoints an r.transition (multiple r.transitions between states are not allowed), elements of an r.transition are extracted using notation *w((s,R),(s',R'))*, *λ((s,R),(s',R'))*. A clock constraint of an ear is called an invariant or an r.state appointed by the ear.

The RCSM automaton is assumed to be **transition-complete**, i.e. for any $(s,R) \in RS$ and for every clock interpretation that fits the region *R*, the disjunction of Boolean formulas *w* for all r.transitions outgoing from *(s,R)* is *true*.

The RCSM automaton is constructed from TCSM using region succession and removing void transitions:

- For any constructed r.state *(s,R)*, *out((s,R))=out(s)*.
- The initial r.state is *($s_{init}$, $R_{init}$)*.
- Construction of RCSM transitions (for *w'* see next point):
  1. For any constructed r.state *(s,R)*, if there is a t.ear $<s,w,\pi,\lambda,s>$ in TCSM, $\lambda=\emptyset$, then construct a progress self-loop (r.ear) from *(s,R)* to *(s,R)*: $<(s,R),w',\emptyset,(s,R)>$.
  2. For any constructed r.state *(s,R)*, if there is a t.ear $<s,w,\pi,\lambda,s>$ in TCSM, $\lambda \neq \emptyset$, then construct a progress r.transition from *(s,R)* to *(s,R')*: $<(s,R),w',\lambda,(s,R')>$, *R'=R[λ:=0]*.

3. For any constructed r.state *(s,R)*, if there is a t.ear $<s,w,\pi,\lambda,s>$ in TCSM, and if $\pi$ agrees with *R*, and if $R' \neq R$ is a successor of *R* with *[λ:=0]*, then construct a successor *(s,R')*, and a progress r.transition $<(s,R),w',\lambda,(s,R')>$.
4. For any constructed r.state *(s,R)* and any t.transition $<s,w,\pi,\lambda,s'>$, $s' \neq s$ in TCSM, if $\pi$ agrees with *R*, then construct an r.state *(s',R')* and a zero-time r.transition $<(s,R),w',\lambda,(s',R')>$ where *R'=R[λ:=0]*.

- The formula *w* in every constructed transition is reduced by *out(s)*: *w'=w\\out(s)* and if the result is ∅ then the transition is rejected.

The rules of RCSM construction guarantee that only reachable part of the graph of the system remains. Some transitions are discarded due to time constraints and some due to output signals are being or not being generated.

Having the succession relation between r.states defined, we may define succession and reachability in RCSM. Succession will be defined for RCSM just as for CSM, taking conjunction of outgoing r.transitions with the output formula.

Given a pair of RCSM automaton r.states belonging to *RS*, $((s,R),(s',R')) \in TS \times TS$, r.state *(s,R')′* is a **region successor** of *(s,R)* iff *form((s,R), (s',R')) \* ω((s,R))* ≠ ∅. The region succession relation is denoted *s rr s'*.

**Region Reachability relation** (denoted **RR**) is a transitive extension of **rr**.

The **Region Reachability Graph (RRG)** of a RCSM automaton is the automaton restricted to r.states reachable from *($s_{init}$,$R_{init}$)*. As the construction of RCSM from TCSM keeps only reachable states, RRG is simply equal to RCSM.

The single step semantics, path semantics and fair path semantics are defined quite similarly to that of CSM and can be found in [8].

## 6 Product of region automata

The disadvantage of "traditional" method of verification is that the whole system of timed automata must be multiplied by every new testing automaton constructed, and then RCSM may be constructed from the product. It is because a multiplication of region automata is not defined for TA (perhaps the reason is interleaving nature of product of TA, which cannot be applied to regions). The product of RCSM is defined below.

Let *P* be a finite set of RCSM automata $P=\{p_i| i=1..n, p_i = <RS_i, out_i, X_i, lab_i, (s_{i,init}, R_{i,init})>\}$. The RCSM automaton $p = <RS, out, X, lab, (s_{init}, R_{init})>$ is a **product** of RCSM automata from *P* (denoted $p \stackrel{def}{=} *_{p_i \in P} p_i$ which means $p_1 * p_2 * ... * p_n$) if sets $\underline{OUT(p_i)}$, *i=1..n* are pairwise disjoint, sets $X_i$, *i=1..n* are pairwise disjoint and:

- $X \stackrel{def}{=} \cup_{p_i \in P} X_i$,
- $RS \subseteq \times_{p_i \in P} RS_i$, ($\times_{p_i \in P} RS_i$ denotes Cartesian product $RS_1 \times RS_2 \times ... \times RS_n$); elements of *RS* (composite states) are pairs of the form:
 $(s,R) \stackrel{def}{=} ((s_{j1}, s_{j2}, ..., s_{jn}),R)$, *R* is a region over *X*;
 *R|i* is a projection of *R* onto a set of clocks $X_i$:

- $R|i = \{I, Y_>, Y_0, Y_{f1}, ..., Y_{fm}\}| i = \{I_i, Y_{>i}, Y_{0i}, Y_{f1i}, ..., Y_{fmi}\}$,
  where $I_i$ is a tuple of integral parts of clocks restricted to $X_i$, all sets indexed by $i$ are conjuncted with $X_i$, then empty sets $Y_{fji}$ are extracted;
- for any $(s,R) \in RS$:
  $out((s,R)) \stackrel{def}{=} \cup_{p_i \in P} out_i(s_{ji})$, $s_{ji} \in RS_i$,
- construction of r.transitions:
  1. Initial r.state $(s_{init}, R_{init}) \in RS$ contains $s_{i,init}$ of all component automata $p_i \in P$:
     $s_{init} \stackrel{def}{=} (s_{1,init}, s_{2,init}, ..., s_{n,init})$,
     $R_{init} \stackrel{def}{=} R[X := 0]$;
  2. For any already constructed r.state $(s,R)$, $s=(s_1,...,s_n)$: take every set of r.transitions of a form $H=\{h_1, ..., h_n\}$ outgoing from $(s_1, R|1), ..., (s_n, R|n)$ in component RCSM automata $p_1...p_n$; for every $H$, $\lambda_H \stackrel{def}{=}_f \cup_{h_i \in H} \lambda_i$.
  3. If a set $H$ contains only progress r.transitions $h_i$, then for one or both region successors of $R$ (one of them is $R[\lambda_H := 0]$, the other one is $R' \neq R$ – a progress successor of $R$ with $[\lambda_H := 0]$,) construct progressive r.transitions:
     - $<(s,R), w, \varnothing, (s, R[\lambda := 0])>$ (a progress r.ear),
     - $<(s,R), w, \varnothing, (s, R'[\lambda := 0])>$ (a progress r.transition with $R' \neq R$), where $w$ is a conjunction of $w_i$ in all r.transitions belonging to $H$, reduced by $out((s,R))$.
  4. If a set $H$ contains at lest one action r.transition, then construct an r.state $s'=(s_1', ..., s_n')$ in which for every action r.transition $h_i$, the state $s_i'$ is a target state of the r.transition $h_i$, and for every progress r.transition $h_i$, $s_i'=s_i$. Then, construct an action r.transition $<(s,R), w, \lambda, (s', R[\lambda_H := 0])>$, where $w$ is a conjunction of $w_i$ of all transitions belonging to $H$ reduced by $out((s,R))$.
- If $w$ in any constructed transition is equal to $\varnothing$, then the transition is rejected (and a state constructed in p.4 as well).

Multiplication of RCSM automata makes it possible to store a set of automata specifying a concurrent system in a form of product RCSM automaton, and multiply it by various testing RCSM automata for verification of desired features. RCSM of system under test once calculated, does not change. Such a procedure cannot be performed if a product of RCSM automata does not exist, in such situation every TCSM test automaton must be multiplied by TCSM system and then RCSM must be calculated.

## 7 Example verification

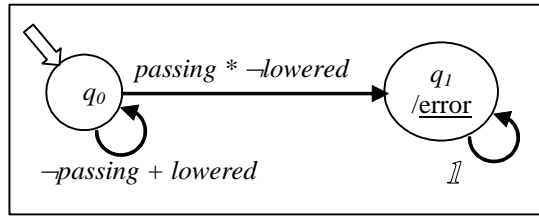

**Fig. 4.** Testing automaton

TCSM may be used for model checking of temporal properties [3]. For example, Alur in [1] defines a correctness condition for a system shown in Fig. 1: if the **TRAIN** is in $s_2'$, the **GATE** should be in $t_2$. This condition can be verified for the system as follows:

- add a signal <u>passing</u> to the t.state $s_1'$,

- add a signal <u>lowered</u> to the t.state $t_2$,
- construct a testing automaton shown in Fig. 4.

The verification consists of the following steps (the case is safety property):
- multiplication of timed automata **TRAIN**, **GATE** and **CONTROLLER**, calculation of system RCSM;
- conversion of testing automaton (Fig. 4) to testing RCSM;
- calculation of RCSM as product of system RCSM with testing RCSM;
- reduction of the product [7,11,12,13];
- observation of the result, the t.state $q_1$ (safety condition) occurs unreachable as needed.

## 8   Conclusions

CSM automata allow modeling of real parallelism, without interleaving in concurrent systems. Timed CSM enhances the formalism to real-time delays .The presented verification technique over TCSM makes it possible to verify concurrent systems with user-specified or automatically generated testing automata. The definition of product of RCSM (which is not defined for Region Automata) allows to store a system under verification in a form ready-to-multiplication with testing automata.

The verification is based on multiplying system RCSM with testing RCSM, reducing the obtained product and searching for reachability of given states (safety properties) or checking the stuttering of given symbols in reduced product (liveness properties).

## 9   Future work

In further research, a verification will be expanded to zone CSM (corresponding to zone automata), which will allow to store state spaces in more compact form.